\documentclass[amsmath,amssymb, 12pt]{revtex4}
\usepackage{graphicx}
\usepackage{dcolumn}
\usepackage{bm}
\usepackage{graphicx}
\usepackage{epsfig}

\def\bea{\begin{eqnarray}}
\def\eea{\end{eqnarray}}

\begin{document}
\title{$\Gamma$-sign in entropic cosmology}
\author{Samuel Lepe}
\altaffiliation{slepe@ucv.cl} \affiliation{Instituto de F\'\i
sica, Facultad de Ciencias, Pontificia Universidad Cat\'olica de
Valpara\'\i so, Avenida Brasil 2950, Valpara\'\i so, Chile.}
\author{Francisco Pe\~na}
\altaffiliation{fcampos@ufro.cl} \affiliation{Departamento de
Ciencias F\'\i sicas, Facultad de Ingenier\'\i a, Ciencias y
Administraci\'on, Universidad de La Frontera, Avda. Francisco
Salazar 01145, Casilla 54-D Temuco, Chile.}

\date{\today}

\begin{abstract}

We study a cosmology by considering entropic issues where the emphasis is
put on the change of sign of the inhomogeneus term associated to an energy
non-conservation equation. We make a bare/effective description of the
equation of state of the cosmic fluid through a bare($\omega $)/effective($%
\omega _{eff}$) conservation equation. In the bare case, where we have a non
conserved equation of state for the cosmic fluid, we describe the change in
the sign of the inhomogeneus term at different times of the cosmic
evolution. We show that by a redefinition of the adiabatic $\omega $%
-parameter we can recover the usual scheme for the cosmic evolution.\ \ In
the effective case we show also that if the evolution is driven by dust or
cosmological constant, the universe evolves on the thermal equilibrium.
Additionally, by incorporating a quantum correction only cosmological
constant can drive an evolution on thermal equilibrium. The future
singularity present in the case when we do not incorporate this correction
is avoided if we do it.
\end{abstract}
\maketitle

\section{ Introduction}

>From the works of E. Verlinde [1] and D. A. Easson et al [2], entropic
gravity is visualized as an alternative to the standard (classical) theory
of gravity described by the Einstein theory. In the first formalism the
space is generated from the termodynamics on a holographic screen. Here the
information is the main ingredient for deriving gravity\ and the input is
the holographic principle; the information is encoded at the boundary. The
second formalism is based on the incorporation of surface terms in the
gravitacional action but gravity is still a fundamental theory. Both schemes
drive to modifications in the usual Friedmann equations of the standard
Friedmann-Lemaitre-Robertson-Walker (FLRW) cosmology and so we can have
another sand where we can inspect. Whether or not these approaches can give
a new scope about gravity is a controversial issue [3].

This letter is organized as follows. In Section II we discuss an effective
scheme ($\omega _{eff}$) for the equation of state (EoS) of the cosmic fluid
through a redefinition of the adiabatic-$\omega $ parameter. In Section III
we describe the sign change in the inhomogeneus term (in the bare EoS for
the cosmic fluid) associated to the interchange of energy between the bulk
and the boundary and we discuss also the absence of future singularities
when a quantum correction is incorporated in the field equations. In Section
IV we summarize the discussed aspects. We will use the units 8$\pi G=$ $c=1$.

------------------------------------------------------------------------

\section{ $\Theta _{eff}$-formulation.}
We discuss the model [4] which is described by the following
modified Friedmann equations

\begin{eqnarray}\label{ec1}
3H^{2} &=&\rho +3\alpha H^{2}+3\beta \dot{H},
\end{eqnarray}
\begin{eqnarray}\label{ec2} \frac{\ddot{a}}{a} &=&\dot{H}+H^{2}=-\frac{1}{6}\left(
3\Theta -2\right) \rho +\alpha H^{2}+\beta \dot{H},
\end{eqnarray}
where the additional terms, $\alpha H^{2}$ and $\beta \dot{H}$ \
are originated on surface terms in the gravitational action and
$\left\{ \alpha ,\beta \right\} $ are two constant parameters
associated to surface curvature where $\alpha =3/2\pi $ and $\beta
=3/4\pi $ and like a generalization this parameters can be bounded
by $\alpha <1$ and $0\leq \beta \lesssim 3/4\pi $ in agreement to
the observational data [2]. The $ \Theta $-quantity is $\Theta
=1+\omega \ $where$\ \omega =p/\rho $ being $p$ the pressure and
$\rho $ the energy density.

We will study two approaches in order to manipulate
(\ref{ec1}-\ref{ec2}): the first is based on making a description
of the model through an effective EoS ($\Theta _{eff}$). The
second approach is based on a non conserved equation for $\rho $.
So, we write the scheme (\ref{ec1}-\ref{ec2}), denoted as
effective scheme, in the form

\begin{eqnarray}\label{ec3}
3H^{2}=\frac{1}{1-\alpha }\left( 1-\frac{3}{2}\beta \Theta \right) \rho ,
\end{eqnarray}

\begin{eqnarray}\label{ec4}
\frac{\ddot{a}}{a}=-qH^{2}=-\left( \frac{3}{2}\Theta _{eff}-1\right) H^{2},
\end{eqnarray}
where $\Theta _{eff}$ is defined through

\begin{eqnarray}\label{ec5}
\Theta _{eff}=\left( 1-\alpha \right) \left( 1-\frac{3}{2}\beta
\Theta \right) ^{-1}\Theta ,
\end{eqnarray}
and by using (\ref{ec4}) the deceleration parameter $q$ is given
by

\begin{eqnarray}\label{ec6}
1+q=\frac{3}{2}\Theta _{eff},
\end{eqnarray}
and the conservation equation becomes

\begin{eqnarray}\label{ec7}
\dot{\rho}+3\Theta _{eff}H\rho =0.
\end{eqnarray}
In the present effective scheme, this equation appears like an
usual energy conservation equation and so, the scheme
(\ref{ec3}-\ref{ec4}) can be saw as one standard FLRW scheme in
the following sense, from (\ref{ec3}-\ref{ec4}) we can obtain the
solution for the Hubble parameter

\begin{eqnarray}\label{ec8}
H\left( t\right) =H_{0}\left[ 1+\frac{3}{2}\Theta
_{eff}H_{0}\left( t-t_{0}\right) \right] ^{-1},
\end{eqnarray}
and this solution it is a one standard given that $\Theta
_{eff}>0\Longrightarrow $ dark matter-evolution or
quintessence-evolution and $\Theta _{eff}<0\Longrightarrow $
phantom-evolution. If we consider early inflation, $\Theta \approx
0$, from (\ref{ec3}-\ref{ec4}) we obtain $H\approx
const.\rightarrow $ $\rho \approx const.$ and the acceleration is

\begin{eqnarray}\label{ec9}
\frac{\ddot{a}}{a}=\frac{1}{3}\left( \frac{1}{1-\alpha }\right)
\rho =const.>0\Longrightarrow \alpha <1,
\end{eqnarray}
so that if we look (\ref{ec3}) we obtain $\beta \Theta <2/3$ and
this constraint for $\beta $\ will be improved later. Particular
cases about bare-effective description $\Theta \left( \omega
\right) -\Theta _{eff}\left( \omega _{eff}\right) $ are:

-cosmological constant

\begin{eqnarray}\label{ec10}
\Theta =0\left( \omega =-1\right) \Longrightarrow \Theta
_{eff}=0\left( \omega _{eff}=-1\right) ,
\end{eqnarray}

and both descriptions, bare and effective, coincide,

-stiff matter

\begin{eqnarray}\label{ec11}
\Theta =2\left( \omega =1\right) \Longrightarrow \Theta
_{eff}=2\left( 1-\alpha \right) \left( 1-3\beta \right)
^{-1}\left( \omega _{eff}=-1+2\left( 1-\alpha \right) \left(
1-3\beta \right) ^{-1}\right) ,
\end{eqnarray}

-dark matter

\begin{eqnarray}\label{ec12}
\Theta =1\left( \omega =0\right) \Longrightarrow \Theta
_{eff}=\left( 1-\alpha \right) \left( 1-\frac{3}{2}\beta \right)
^{-1}\left( \omega _{eff}=-1+\left( 1-\alpha \right) \left(
1-\frac{3}{2}\beta \right) ^{-1}\right) ,
\end{eqnarray}

-phantom dark energy

\begin{eqnarray}\label{ec13}
\Theta <0\left( \omega <-1\right) \Longrightarrow \Theta
_{eff}<0\left( \omega _{eff}<-1\right) ,
\end{eqnarray}

and in this case we are also coincidence in both schemes.

By using (\ref{ec6}) we can fit the q-parameter in the following
form

\begin{eqnarray}\label{ec14}
\Theta _{eff}=1\text{ }\left( q=1/2\right) \text{\ \ }and\text{ \
}\Theta =1 \text{ }\Longrightarrow \alpha =\frac{3}{2}\beta ,
\end{eqnarray}
so that for dark matter we have $\omega _{eff}=\omega =0$. Thus, we can write

\begin{eqnarray}\label{ec15}
\Theta _{eff}=\left( 1-\alpha \right) \left( 1-\alpha \Theta
\right) ^{-1}\Theta ,
\end{eqnarray}
and we have one parameter ($\alpha $) for fixing. Now, by considering $%
\Theta =2\left( \omega =1\right) $, i.e. stiff matter, we have

\begin{eqnarray}\label{ec16}
\Theta _{eff}=2\left( 1-\alpha \right) \left( 1-2\alpha \right)
^{-1}\rightarrow \omega _{eff}=-1+\frac{1-\alpha }{1/2-\alpha },
\end{eqnarray}
and $1/2<\alpha <1\Longrightarrow \omega _{eff}<-1,$ i.e., something like
effective phantom stiff matter! We solve this problem by doing $\alpha <1/2$%
, and in this case we have a quintessence-evolution driven for stiff matter.
If $0<\alpha <<1/2$ we have $\omega _{eff}$\ $\approx 1$.\ We note that $%
\alpha <1/2\Longrightarrow \beta <1/3$ (see (\ref{ec14})). So, the
present effective scheme result to be on standard... and nothing
new under the sun!, that is,\ we obtain the same as the usual
FLRW-scheme by changing $\Theta \rightarrow \Theta _{eff}$.

\section {$\Gamma $-formulation.}
We come back to (\ref{ec7}) in order to obtain

\begin{eqnarray}\label{ec17}
\dot{\rho}+3\Theta H\rho =\Gamma ,
\end{eqnarray}
where $\Gamma $ is the amount of energy non-conservation which is given by \
\ \ \ \ \ \ \ \ \ \ \ \ \ \ \ \ \ \ \ \ \ \

\begin{eqnarray}\label{ec18}
\Gamma =3\left( \Theta -\Theta _{eff}\right) H\rho =3\alpha \Theta
\left( \frac{1-\Theta }{1-\alpha \Theta }\right) H\rho ,
\end{eqnarray}
where we have used (\ref{ec15}). This quantity, $\Gamma $,
represents the interchange of energy between the bulk (the
universe) and the boundary (Hubble horizon). At early times,
$\Theta \approx 0\longleftrightarrow H\approx const.$and $\rho
\approx const.$(exponential inflation), so that we have $\Gamma
\approx 0$ and the same occours if we have dark matter $\left(
\Theta =1\right) $. In the case of stiff matter $\left( \Theta
=2\right) $ or radiation $\left( \Theta =4/3\right) $ we obtain
$\Gamma <0$ and the same
occours at late times if we have the possibility of phantom dark energy $%
\left( \Theta <0\right) $, and we note that $\Gamma >0$ if we are in the
quintessence-zone $\left( 0<\Theta <1\right) $. In the case of cosmological
constant $\left( \Theta =0\right) $ we have $\Gamma =0$.

\ By using the eqs. (\ref{ec15}), (\ref{ec8}) and (\ref{ec1}), the
expression (\ref{ec18}) can be writen in the equivalent forms

\begin{eqnarray}\label{ec19}
\Gamma \left( \Theta ,\alpha ;t\right) &=&9H_{0}^{3}\alpha \left( 1-\alpha
\right) \frac{\left( 1-\Theta \right) \Theta }{\left( 1-\alpha \Theta
\right) ^{2}}\left[ 1+\frac{3}{2}\frac{\left( 1-\alpha \right) \Theta }{%
\left( 1-\alpha \Theta \right) }H_{0}\left( t-t_{0}\right) \right] ^{-3},
\end{eqnarray}
\begin{eqnarray}\label{ec20}
&=&\frac{8}{3}H_{0}^{3}\frac{\alpha }{\left( 1-\alpha \right) ^{2}}\frac{%
\left( 1-\Theta \right) \left( 1-\alpha \Theta \right) }{\Theta ^{2}}\left[
\frac{2}{3}\frac{\left( 1-\alpha \Theta \right) }{\left( 1-\alpha \right)
\Theta }+H_{0}\left( t-t_{0}\right) \right] ^{-3},
\end{eqnarray}
\begin{eqnarray}\label{ec21}
&=&-\frac{8}{3}\frac{\alpha }{\left( 1-\alpha \right)
^{2}}\frac{\left( 1-\Theta \right) \left( 1-\alpha \Theta \right)
}{\Theta ^{2}}\left( t_{s}-t\right) ^{-3},
\end{eqnarray}
where

\begin{eqnarray}\label{ec22}
t_{s}=t_{0}-\frac{2}{3\Theta _{eff}}H_{0}^{-1}=t_{0}-\frac{2\left(
1-\alpha \Theta \right) }{3\left( 1-\alpha \right) \Theta
}H_{0}^{-1},
\end{eqnarray}
and we can see that $\Gamma \left( 1,\alpha ;t\right) =\Gamma
\left( 0,\alpha ;t\right) =\Gamma \left( \alpha ^{-1},\alpha
;t\right) =0$ and we have a future singularity for $\Theta <0$
(this singularity will be removed by the inclusion of quantum
corrections on (\ref{ec1}- \ref{ec2}), see later) and a past
singularity if $\Theta >\alpha ^{-1}$. If $0<\alpha
<1/2\Longrightarrow \Theta
>2$, i.e., $\omega >1$ (super stiff matter), but the observational
data discards this past singularity. So,%
\begin{eqnarray}\label{ec23}
\begin{array}{c}
\Gamma \left( \Theta >1,\alpha ;t\right) <0\Longleftrightarrow \omega >0%
\text{ }\left( beyond\text{ }dust-zone\right) , \\
\Gamma \left( 1,\alpha ;t\right) =0\Longleftrightarrow \omega =0\text{ }%
\left( dust\right) \\
\Gamma \left( 0<\Theta <1,\alpha ;t\right) >0\Longleftrightarrow -1<\omega <0%
\text{ }\left( qu\mathit{in}tessence-zone\right) , \\
\Gamma \left( 0,\alpha ;t\right) =0\Longleftrightarrow \omega =-1\text{ }%
\left( \cos mo\log ical\text{ }cons\tan t\right) \\
\Gamma \left( \Theta <0,\alpha ;t\right) <0\Longleftrightarrow \omega <-1%
\text{ }\left( phantom-zone\right) ,%
\end{array}
\end{eqnarray}
and the sign of $\Gamma $ it`s not always the same during the cosmic
evolution. So, the evolution is developed out the thermal equilibrium at
exception when we have dust or cosmological constant.

We come back now to (\ref{ec4}). By using (\ref{ec18}) we write
the following expression for the acceleration

\begin{eqnarray}\label{ec24}
\frac{\ddot{a}}{a}=\frac{3}{2}H^{2}\left( \frac{2}{3}+\frac{\Gamma
}{3H\rho } -\Theta \right) ,
\end{eqnarray}
and we note that

\begin{eqnarray}\label{ec25}
\Theta =\frac{2}{3}\longleftrightarrow \omega
=-\frac{1}{3}\Longrightarrow \frac{\ddot{a}}{a}=\left(
\frac{H}{2\rho }\right) \Gamma \text{ \ \ }and \text{ \ \ }\Gamma
=\left( \frac{\alpha }{3/2-\alpha }\right) H\rho >0.
\end{eqnarray}
In standard cosmology, $\Theta =2/3\longleftrightarrow \omega =-1/3$
correspond to a curvature fluid (string gas) and in that case we have $\ddot{%
a}=0$. Nevertheless, in the present case we have $\ddot{a}>0$. What happen
today? From WMAP 5-7 [5]: $1+\omega \left( 0\right) =\Theta \left( 0\right)
\approx -0.10\pm 0.14$ (for $\omega -$time independent) so that we define $%
\Theta ^{+}\left( 0\right) =0.04$ and $\Theta ^{-}\left( 0\right)
=-0.24$. Thus, in accord to (\ref{ec18})

\begin{eqnarray}\label{ec26}
\Gamma \left( 0\right) >0\longleftrightarrow \Theta \left(
0\right) =\Theta ^{+}\left( 0\right) \text{ \ \ }and\text{ \ \
}\Gamma \left( 0\right) <0\longleftrightarrow \Theta \left(
0\right) =\Theta ^{-}\left( 0\right) ,
\end{eqnarray}
and $\Gamma \left( 0\right) =0$ if the evolution is one standard driven by
the cosmological constant, and in this case we do not notice today the
"presence" of $\Gamma $ (the same happened when $\Theta =1$, evolution
driven by dust). So, we can say that $\Gamma \left( 0\right) \approx 0$ and
today the cosmic fluid is one conserved (delicate thermal equilibrium
between the bulk and the boundary today?). We note that in the quintessence
zone $\Theta =1\left( \omega =0\right) $ and $\Theta =0\left( \omega
=-1\right) $ we have energy transference from the boundary to the bulk ($%
\Gamma >0$). If the future is driven for phantom dark energy, then we have $%
\Gamma <0$ plus a singularity and in this case we will have energy going
from the bulk to the boundary and the temperature of the boundary ($%
T_{H}=H/2\pi $) will increase and we ask, what will happen when
the singularity is nearby?, super hot boundary ($T_{H}\rightarrow
\infty $) and bulk frozen ($T\rightarrow 0$)? We inspect now the
phantom zone ($\Theta <0\rightarrow \Gamma <0$). The Hubble
temperature is given by $T_{H}\left( t\right) =H\left( t\right)
/2\pi $ and we do the following exercise (in order to have a
feeling): by using (\ref{ec8}) with $\Theta _{eff}<0$ (phantom
zone) we obtain

\begin{eqnarray}\label{ec27}
H\left( a\right) =H_{0}\left( \frac{a}{a_{0}}\right) ^{3\left\vert
\Theta _{eff}\right\vert /2},
\end{eqnarray}
so that the boundary temperature is

\begin{eqnarray}\label{ec28}
T_{H}\left( a\right) =T_{H}\left( 0\right) \left(
\frac{a}{a_{0}}\right) ^{3\left\vert \Theta _{eff}\right\vert /2}.
\end{eqnarray}
For radiation ($\Theta =4/3$ and $\Theta _{eff}=\left( 4/3\right) \left(
1-\alpha \right) /\left( 1-4\alpha /3\right) $) we have

\begin{eqnarray}\label{ec29}
T_{r}\left( a\right) =T_{r}\left( 0\right) \left(
\frac{a}{a_{0}}\right) ^{-\left( 1-\alpha \right) /\left(
1-4\alpha /3\right) },
\end{eqnarray}
and we do now $T_{H}\left( \bar{a}\right) =T_{r}\left( \bar{a}\right) $
(equilibrium) so that

\begin{eqnarray}\label{ec30}
\bar{a}=a_{0}\left( \frac{T_{r}\left( 0\right) }{T_{H}\left(
0\right) } \right) ^{1/\left( \Delta +3\left\vert \Theta
_{eff}\right\vert /2\right) },
\end{eqnarray}
where $\Delta =\left( 1-\alpha \right) /\left( 1-4\alpha /3\right) $. Today $%
T_{r}\left( 0\right) \sim 3K$ and $T_{H}\left( 0\right) \sim 0\left(
10^{-30}\right) K$ and in this case we have thermal equilibrium at $\bar{a}%
\sim 10^{12}a_{0}$! We note that $T_{r}\left( \bar{a}\right) \sim
10^{-12}T_{r}\left( 0\right) $ and $T_{H}\left( \bar{a}\right) \sim
10^{18}T_{H}\left( 0\right) \sim 10^{-12}K$. The time ($t_{eq}$) at which we
obtain this thermal equilibrium is given by

\begin{eqnarray}\label{ec31}
t_{eq}-t_{0}=\left( t_{s}-t_{0}\right) \left( 1-\frac{T_{H}\left( 0\right) }{%
T_{CMB}\left( 0\right) }\right) ^{1/\Omega },
\end{eqnarray}
where $\Omega =1+2/3\left\vert \Theta _{eff}\right\vert $. So, we have $%
t_{eq}\sim t_{s}$, and we have thermal equilibrium very near to
the singularity.

Now, by adding the quantum corrections $3\gamma H^{4}$ in
(\ref{ec1}) and $\gamma H^{4}$ in (\ref{ec2}) [4,6] it is
straightforward to obtain

\begin{eqnarray}\label{ec32}
3H^{2}=\frac{1}{1-\alpha }\left( 1-\frac{3}{2}\beta \Theta \right)
\rho +3 \frac{\gamma }{1-\alpha }H^{4},
\end{eqnarray}
and

\begin{eqnarray}\label{ec33}
\dot{H}+H^{2}=-\left( \frac{3}{2}\tilde{\Theta}-1\right) H^{2}
\end{eqnarray}
where

\begin{eqnarray}\label{ec34}
\tilde{\Theta}=\Theta _{eff}\left( 1-\frac{\gamma }{1-\alpha
}H^{2}\right) ,
\end{eqnarray}
and $\Theta _{eff}$\ \ is given by (\ref{ec5}).\ So, the new
effective scheme becomes

\begin{eqnarray}\label{ec35}
\dot{\rho}+3\tilde{\Theta}H\rho =0,
\end{eqnarray}
and the new bare scheme is

\begin{eqnarray}\label{ec36}
\dot{\rho}+3\Theta H\rho =\tilde{\Gamma},
\end{eqnarray}
where

\begin{eqnarray}\label{ec37}
\tilde{\Gamma}=3\Theta \left[ 1-\left( 1-\alpha \right) \left(
1-\frac{3}{2} \beta \Theta \right) ^{-1}\left( 1-\frac{\gamma
}{1-\alpha }H^{2}\right) \right] H\rho .
\end{eqnarray}
If we consider $\alpha =\left( 3/2\right) \beta $ we obtain

\begin{eqnarray}\label{ec38}
\tilde{\Gamma}=3\alpha \Theta \left( \frac{1-\Theta }{1-\alpha
\Theta } \right) \left[ 1+\frac{\gamma }{\alpha
}\frac{H^{2}}{1-\Theta }\right] H\rho
\end{eqnarray}
and at difference of $\Gamma \left( \Theta ;t\right) $
(\ref{ec23}) we can see that only $\Theta =0\left( \omega
=-1\right) $ implies $\tilde{\Gamma}=0$. So, when the evolution is
driven by dust we have not thermal equilibrium if we consider the
quantum correction.

Now, from Eqs. (\ref{ec32}-\ref{ec34}) it is easy to obtain the
following implicit solution for the Hubble parameter

\begin{eqnarray}\label{ec39}
\frac{3}{2}\Theta _{eff}\left( t-t_{0}\right)
=\frac{1}{H}-\frac{1}{H_{0}}+ \frac{1}{2\delta }\ln \left(
\frac{H-\delta }{H_{0}-\delta }\right) \left( \frac{H_{0}+\delta
}{H+\delta }\right) ,
\end{eqnarray}
where we are defined

\begin{eqnarray}\label{ec40}
\delta =\sqrt{\frac{\left( 1-\alpha \right) }{\gamma }}.
\end{eqnarray}
If we consider $\Theta _{eff}<0$, Eq. (\ref{ec40}) can be written
in the form

\begin{eqnarray}\label{ec41}
\frac{1}{H}-\frac{3}{2}\left\vert \Theta _{eff}\right\vert \left(
t_{s}-t_{0}\right) =\frac{1}{2\delta }\ln \left( \frac{1+\delta
/H}{1-\delta /H}\right) \left( \frac{1-\delta /H_{0}}{1+\delta
/H_{0}}\right) ,
\end{eqnarray}
where $t_{s}$ is given in (\ref{ec22}) by doing $\Theta
_{eff}=-\left\vert \Theta _{eff}\right\vert $, and it is easy to
verify when $t=t_{s}$ there is not future singularity for $H$ (in
fact, when $t=t_{s}$\ Eq. (\ref{ec41}) is satisfied only for
$H=H_{s}<H_{0}$). Finally, in accord to Eq. (\ref{ec32}) we can
visualize the auto-accelerated solution $\rho =0\Longrightarrow
H=\sqrt{\left( 1-\alpha \right) /\gamma }$ and in this case
$\tilde{\Gamma}=0$ too and under phantom evolution we have
$\tilde{\Gamma}\neq 0$. \ \ \ \

\section {Final remarks}

We have studied the entropic model given in (\ref{ec1}-\ref{ec2})
by doing a bare/effective description of the equation of state for
the cosmic fluid. The effective description have showed to be an
usual, that is, only by redefining the parameter of the equation
of state we find a standard FLRW cosmology. In the bare case, we
find sign changes in the term which accounts the state of thermal
equilibrium, and only when the evolution is driven by dust or
cosmological constant we have thermal equilibrium\textbf{. }We
have showed that during a phantom evolution it is possible to
reach the thermal equilibrium between the bulk (radiation) and the
boundary (Hubble horizon) in the nearby of the singularity.
Finally, by adding a quantum correction to the modified Friedmann
`equations only cosmological constant can drive the universe on
thermal equilibrium and the future singularity, which it is
present in absence of the quantum corrections, is avoided. So,
under the scope of the entropic cosmology is it possible to have a
phantom-free evolution.

\section{acknowledgements}

This work was supported by FONDECYT Grant\ No. 1110076 (SL),
VRIEA-DI-037.419/2012 (SL), Vice Rector\'{\i}a de
Investigaci\'{o}n y Estudios Avanzados, Pontificia Universidad
Cat\'{o}lica de Valpara\'{\i}so (SL) and DIUFRO DI12-0006,
Direcci\'{o}n de Investigaci\'{o}n y Desarrollo, Universidad de La
Frontera (FP). One of us (SL) acknowlwdge the hospitality of
Departamento de Ciencias F\'{\i}sicas de la Universidad de La
Frontera where part of this work was made.

\textbf{References}

[1] E. P. Verlinde, JHEP \textbf{1104} (2011) 029.

[2] D. A. Easson, P. H. Frampton and G. F. Smoot, Phys. Lett. \textbf{B}
\textbf{696}: 273 - 277, 2011.

[3] A. Kobakhidze, arXiv: 1108.4161 and Phys. Rev. D \textbf{83} (2011)
021502,

\ \ \ \ Shan Gao, Entropy \textbf{13} (2011) 936-948,

\ \ \  M. Chaichian, M. Oksanen and A. Tureanu, Phy. Lett. B \textbf{702}
(2011) 419-421.

[4] T. S. Koivisto, D. F. Mota and M. Zumalacarregui, JCAP \textbf{1102}
(2011) 027.

[5] E. Komatsu et al, Astrophys. J. Suppl. \textbf{192}:18, 2011.

[6] Yi-Fu Cai and E. M. Saridakis, Phys. Lett. B \textbf{697} (2011) 280-287.

\end{document}